# Science, Art and Geometrical Imagination

Jean-Pierre Luminet
Laboratoire Univers et Théories, CNRS-UMR8102,
Observatoire de Paris, 92195 Meudon Cedex (France)

Abstract

From the geocentric, closed world model of Antiquity to the wraparound universe models of relativistic cosmology, the parallel history of space representations in science and art illustrates the fundamental role of geometric imagination in innovative findings. Through the analysis of works of various artists and scientists like Plato, Dürer, Kepler, Escher, Grisey or the present author, it is shown how the process of creation in science and in the arts rests on aesthetical principles such as symmetry, regular polyhedra, laws of harmonic proportion, tessellations, group theory, etc., as well as beauty, conciseness and emotional approach of the world.

## 1. Symmetry and Polyhedra in Art and Cosmology

The concept of universal harmony and a geometric view of the world are particularly relevant to the study of the sky [1]. In the sixth century BC around the Mediterranean the Pythagoreans first developed a cosmic theory based on proportion, numbers and resonances like notes in the scale. Such ideas became widely accepted and Plato, in the fourth century BC, adopted the word *kosmos* – which had hitherto been used to describe women's apparel (cf. cosmetics), ornamentation, physical and moral attractiveness, order, truth, etc. – to denote the earth and the stars. The cosmos was also the primary subject of poetry, the poetic world being a reflection of the real world, and became synonymous with the idea of a vast and majestic universe governed by the principles of aesthetics, order and harmony.

Plato's hypothesis was of an organised cosmos whose laws could be deciphered, explained in geometric terms [2]. He knew that the motion of the planets, as well as that of the sun and stars, was not random but obeyed certain laws and was therefore predictable ; that was the starting point for the application of geometry to an understanding of the workings of the cosmos.

Together with his pupil Aristotle [3], they introduced the fundamental distinction between sublunary and superlunary worlds (fig.1). The former is made from the four basic elements, earth, water, air and fire ; it is the world of change and corruption : creatures and things are born, grow, wear out and die. The latter, comprising the celestial spheres, heavens and the firmament, is the world of perfection, eternal and unchanging.

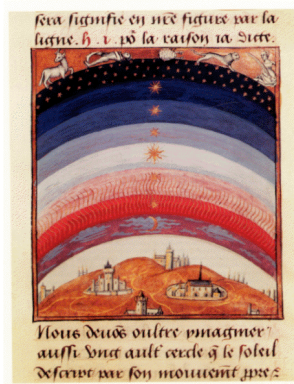

**Fig. 1**. As shown in this beautifully painted medieval manuscript, Aristotle saw the universe as being divided into a perfect, « superlunary » region and an imperfect, « sublunary » region. This distinction would continue to be made until the early 16[th] century.

Plato then developed his ideas on representing the « two worlds » in geometric terms. The perfect superlunary world is organised with spheres – a sphere being « shape which incorportates all other shapes », perfect, harmonious, symmetrical and uniform. The sublunary world is far less harmonious than the superlunary world, consequently each of the four elements comprising it must be represented by a shape somewhat less symmetrical, less pleasing and perfect than a sphere, i.e. one of five « Platonic solids », nowadays known as regular polyhedra (fig. 2). Plato thought carefully about the relationship between each element and its representative shape ; for example, a cube is the most difficult shape to move, so it is associated with earth, the heaviest element ; an icosahedron has more sides than any other Platonic solids (five triangles meet at each point), giving it a virtually fluid structure which is most clearly associated with water, and so on. To Plato each perfect solid represented the essence of its corresponding element so, when its contemporary Theaetetus pointed out to him that the dodecahedron was the fifth regular polyhedron, Plato postulated a fifth element in order to unify his geometric model of the cosmos. In the Middle Ages the fifth element was to be known as « quintessence », but in *Epinomis*, a work published after *Timaeus* and attributed to Plato, it is called *aither*, meaning « eternal flux ».

Of the five regular polyhedra the dodecahedron is the nearest to a sphere, the symbol of celestial perfection. As Plato put it, « God used this solid for the whole universe, embroidering figures on it. ». Since then, the dodecahedron has been charged with a heavy symbolism. The number five, associated to the number of sides of its polygonal faces, plays a particular role in occultism: the pentacle or five-pointed star, inside of which are inscribed letters, words, and signs, is supposed to translate a universal structure. The number 12, which is the number of faces, puts it naturally in correspondence with the 12 signs of the zodiac, the 12 months of the year, the 12 apostles, and so forth. Bronze artifacts of gallo-roman origin have been found in dodecahedral form.

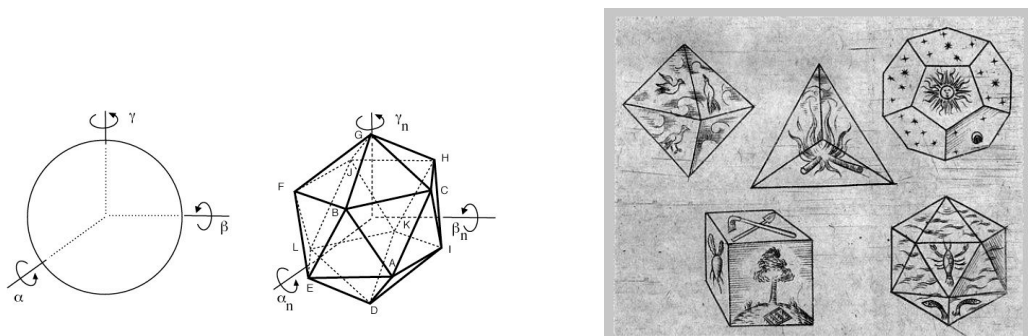

**Fig. 2.** *Left :* The group of rotations. A sphere keeps the same shape whichever way it is rotated about its three axes. The corresponding group of rotation is said to be « continuous ». A regular polyhedron, on the other hand, only retains its shape of the angle through which it is turned takes its values in a discrete set. The corresponding group of symmetry, which is a subset of the full group of rotations, is said to be discontinuous. *Right :* Euclid first demonstrated that there are only five regular polyhedra in 3-dimensional space. As depicted here in Kepler's *Harmony of the World* (1619), for the ancient Greek philosophers earth was associated to a cube, water to an icosahedron, air to an octahedron, fire to a tetrahedron and etheral « fifth » essence to a dodecahedron.

Plato and Aristotles's concepts of the cosmos enabled them to construct a compact, closed model of the world, based on interlocking spheres representing a concentric arrangement of stars and planets revolving circularly around the earth – the latter, being the heaviest element, stays immobile at the center of the cosmos, which is therefore geocentric (fig. 3).

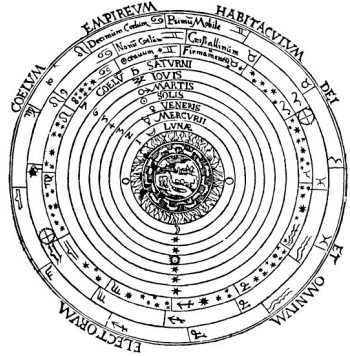

**Fig. 3.** The geocentric, closed universe model according to Aristotle, seen through the eyes of Medieval Christianity, with modifications such as the empyrean, the abode of God and the elects. From Peter Apian, *Cosmographicus Liber*, 1524.

Platonic solids were central to the way the world was represented in succeeding centuries. In the Renaissance, artists such as Paolo Uccello (1397 – 1475) and Piero della Francesca (1415 ?-1492) were fascinated by them. With his precise, analytical mind Uccello applied a scientific method to depict objects in three-dimensional space. In particular, some of his studies of the perspective foreshortening of the torus are preserved, and one standard display of drawing skill was his depictions of the *mazzochio*.

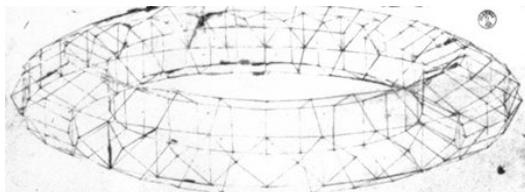 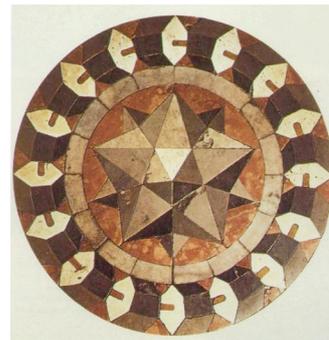

**Fig. 4.** *Left :* Uccello's fresco *The Flood* (1448) shows a figure wearing a *mazzocchio* – a hat in the form of a torus, worn by wealthy men in 15th century Florence. Here is a preparatory drawing in which Uccello shows the form of a polyhedral torus in careful perspective. *Right :* Generally attributed to Uccello, this marble inlay features a small stellated dodecahedron, located in the floor of the Basilica of St. Mark in Venice. It is remarkable, for this would be two hundred years before Kepler's 1619 mathematical description of this same polyhedron.

According to Giorgio Vasari, Piero della Franceca was the best geometer of his time. Three treatises written by Piero are known to modern mathematicians, including *Short Book on the Five Regular Solids (Libellus de Quinque Corporibus Regularibus)*. The subjects covered in these writings include arithmetic, algebra, geometry and innovative work in both solid geometry and perspective. Much of Piero's work was later absorbed into the writings of others, notably the Italian humanist and mathematician Luca Pacioli (1445 ?-1514 ?), a member of the minor order of monks and a professor of theology.

In his *Divine Proportion* of 1509, Pacioli used Platonic solids to define the laws of harmonic proportion, applicable to music, architecture, calligraphy and other arts. The notion of harmonic proportion was linked to the celebrated golden ratio [4] (in mathematics and the arts, two quantities are in the golden ratio if the ratio of the sum of the quantities to the larger one equals the ratio of the larger one to the smaller.) Containing illustrations of regular solids by Leonardo da Vinci, Pacioli's longtime friend and collaborator, *De Divina Proportione* had a major influence on generations of artists and architects alike (fig. 5).

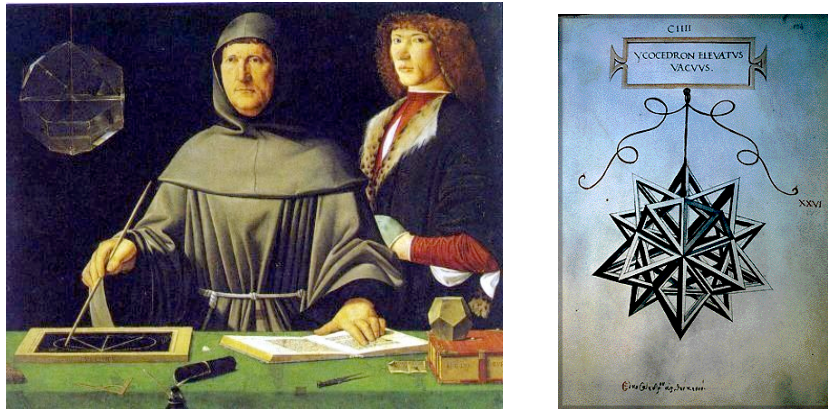

**Fig. 5.** *Left :* This painting of Luca Pacioli, attributed to Jacopo de Barbari (1495), depicts a table filled with geometrical tools: slate, chalk, compass, a dodecahedron model. A rhombicuoctahedron half-filled with water is suspended from the ceiling. Pacioli is demonstrating a theorem by Euclid to a student. *Right : Divine Proportion* was the last published work of Pacioli. A former pupil of Piero della Francesca, Pacioli asked Leonardo da Vinci to draw the diagrams for his book. Here, a stellated icosahedron.

Pacioli actually advocated the Vitruvian system of rational proportions. The source of classical knowledge on architecture was Vitruvius (I$^{rst}$ century BC) and his *Ten Books on Architecture*. The Vitruvian ideal, shared by all Renaissance humanists, was that an architect must be what is now termed a Renaissance man, a « man of letters, a skillful draftsman, a mathematician, familiar wih scientific inquiries, a diligent student of Philosophy, acquainted with music, not ignorant of medicine, learned in the responses of juriconsults, familiar with astronomy and astronomical calculations. »[5]

Vitruvius had emphasized the human dimension of cosmic unity by describing it in terms of the microcosm. His standard of symmetry and proportion was the human body, the famous Vitruvian man that intrigued the artists and architects of the Renaissance, to begin with Leonardo da Vinci's *Homo Universalis* (fig. 6 left), the theoretical work of Leon Battista Alberti about painting, and many others who used the proportions of the human body to improve the appearance and function of architecture, believing these proportions to be aesthetically pleasing. As the golden ratio can be deduced from the proportions of the regular pentagon, it is also structural to the construction of a dodecahedron, as it can be seen in a painting of Nicolas Neufchatel (fig. 6 right).

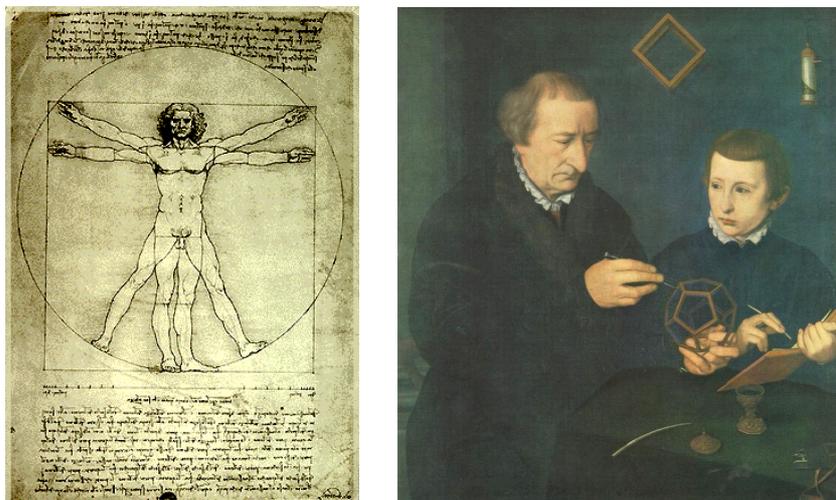

**Fig. 6.** *Left :* Leonardo da Vinci, *Homo Universalis*, inspired from Vitruve. *Right :* Nicolas Neufchatel (1527-1590), *Portrait of Johannes Neudörfer and his Son*, Oil on canvas, Alte Pinakothek, Munich.

In 1525 Albrecht Dürer (1471-1528) published his treatise *Four Books on Measurement* at Nuremberg (cited later by Galileo and Kepler). The fourth book deals with the geometry of platonic solids, Archimedean semi-regular solids, as well as several of his own invention, and is illustrated with wonderful drawings.

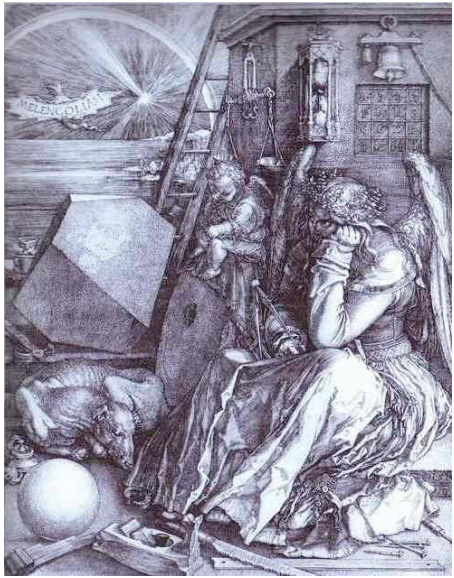

**Fig. 7.** Melancholy, or the Spirit of Man in Search of the Secret of the Universe. This Dürer's etching, dating from 1514 according to the numbers in the square in the top right corner, depicts man contemplating the nature of the world in the state of melancholy, which in medieval times was associated with black bile and the planet Saturn [6]. The winged man prefigures Johannes Kepler's interrogations as he calculates how to express the underlying harmony of the cosmos using spheres and polyhedra. The bright light in the sky is the great comet that was observed in the winter of 1513-1514. As it shines on the scales (depicting the astronomical sign Libra) it symbolises the end of an earthly cycle, if not the end of time itself. The ladder with seven rungs represents the belief held by the Byzantines that the world would not exist for more than seven thousand years. It is the end of the Middle Ages; Dürer, a contemporary of Copernicus, is to be one of the prime movers of the Renaissance.

Wenzel Jamnitzer (1508-1585) was a follower of Dürer. His *Perspectiva Corporum Regularum*, published in 1568, depicts subtle geometric variations of regular polyhedra, including stellations. Each regular solid, associated to each element, is represented by four plates each containing six different aspects of the associated polyhedron. His book was very influential and belonged to the personal libraries of Tycho Brahe and Kepler.

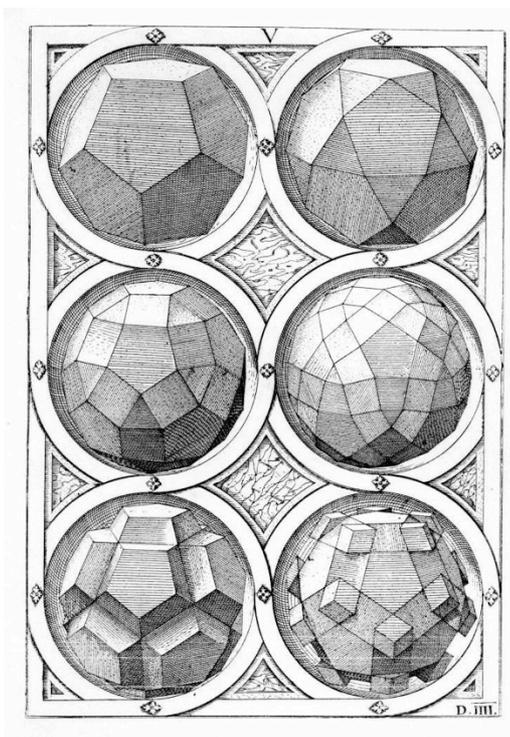

**Fig. 8.** Variations on the dodecahedron. To each regular polyhedron Jamnizer associated a vowel : A for the tetrahedron, E for the octahedron, I for the cube, O for the icosahedron and U (written V) for the dodecahedron. Such a correspondence is linked to an old tradition which assigned an esoteric meaning to the letters of the alphabet, and in particular to the vowels. The meaning of the couple Alpha-Omega is well-known to design the beginning and the end. Jamnitzer chose the A for the tetrahedron, namely the origin for the most simple polyhedron, and the U (equivalent to the greek Omega) for the dodecahedron, namely the most complex polyhedron.
Plate from Wenzel Jamnitzer, *Perspectiva Corporum Regularum*, Nuremberg, 1568.

As a humanist, Nicolaus Copernicus (1473-1543) was interested in medicine, politics, the art of warfare, painting, etc. as well as in science. In this sense he was the first Renaissance scientist [7]. During a 10 years stay in Italy at the end of the 15$^{th}$ century he came under powerful influence of Italian artists and thinkers, who were beginning to find solutions to the problem of representing three-dimensional space – in other words perspective. These were incorporated into the Copernican astronomical system, designed for improving the errors in the positioning of Venus and Mars as well as discrepancies in Ptolemy's predictions of lunar eclipses, etc. But the deep motivation of Copernicus was aesthetic, as he declared in his *De Revolutionibus Orbium Coelestium,* published in Nuremberg en 1543 : « In this arrangement, therefore, we discover a marvellous symmery of the universe, and an established harmonious linkage between the motion of the spheres and their size, such as can be found in no other way. »[8] For this, Copernicus revived Aristarchus' heliocentric model of the solar system, based on geokinetics. Copernicus understood that our viewpoint of the cosmos is constantly changing. Just as Brunelleschi's window (*tavoletta*) enabled him to discover the laws of perspective, so the earth allows us to discover the laws of the cosmos ; but ours is a moving window, and it is that movement which explains the apparent motion of the planets. The application of perspective to man's view of the sky was therefore fundamental to the development of the Copernican system in determining the structure of the cosmos and the relationship between its constituents. Moreover, the fact that the earth moves gives us the opportunity (unavailable to believers in the Ptolemaic system) to measure parallaxes and therefore to calculate the distances of the stars. From then on perspective would remain an integral part of astronomy, governing every attempt to conceptualise and to explore the universe.

The professor of mathematics at the University of Tübingen, Michael Maestlin (1550-1631), was one of the first to accept and teach the heliocentric Copernican view. Among his students was Johannes Kepler. Maestlin corresponded with him frequently and played a sizable part in his adoption of the Copernican system. Galileo Galilei's adoption of heliocentrism is also attributed to Maestlin. Note that the first known calculation of the inverse golden ratio as a decimal of « about 0.6180340 » was written in 1597 by Maestlin in a letter to Kepler.

His contemporary, the great Danish astronomer Tycho Brahe (1546-1601), was unable to bring himself to remove the earth from the centre of the cosmos. Knowing that a strictly geocentric model was no more able to account for the motion of the planets, he proposed in 1588 a combined system, which enjoyed considerable popularity : the earth remained fixed at the centre, the sun, the moon and the fixed stars revolving around it, but the other five planets and the comets revolved around the sun. Nevertheless, trained in the tradition of humanist learning, Tycho Brahe was fascinated by celestial harmony [9].

In 1576 Brahe's patron Frederick II, king of Denmark, granted him permission to construct a magnificent house and observatory, a temple to astronomy called Uraniborg, on the island of Hven. Tycho was thoroughly familiar with Vitruvius' *Books on Architecture*. He was also strongly impressed by the more recent architecture he had seen on his travels in northern Italy, and owned the magnificent illustrated editions of his contemporaries Serlio and Palladio. Thus Brahe designed the building according to the theory of divine proportion, which had been applied to architecture principally by Palladio (1518-1580). Palladio's *Four Books of Architecture*, published in Venice in 1570, was the most thourough possible application of microcosmic proportion and symmetry to architecture. His system of proportions included the use of harmonic ratios derived from musical theory. Like the human body, Palladio's buildings were symmetrical to a central axis, with single elements along the axes – like the nose or mouth of the body – and lateral elements in pairs – like the eyes, ears or arms of the body. Fundamental to the microcosmic plan of these structures was the Vitruvian figure of human proportion within homocentric squares and circles. In its most simple and monumental form, the unit of microcosm and macrocosm, the central role of humanity in the universe, and the merging of God's spirit with the world's mathematical structure could be expressed in a building that was at once a work of art, an emblem, and a human dwelling.

The Palladian ideal, illustrated by the Villa Rotonda in Vicenze, was that a country house was really a small city, each being a reflection of the cosmos (fig. 9). For Palladio, architecture imitated the geometrical sense of order that constituted the hidden framework of the universe. If an astronomer like Tycho had achieved a certain understanding of the universe, he could express that understanding through architecture. In the case of Tycho's castle on Hven island, it would also be a temple for the worship of Apollo, Mercury and the muses (fig. 10).

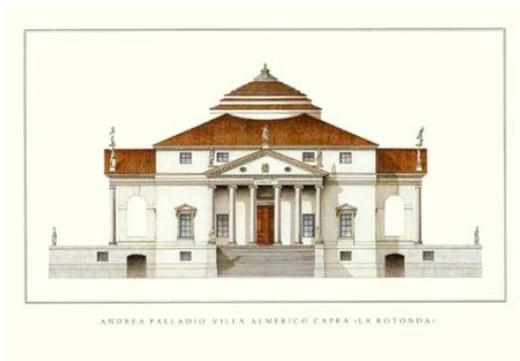 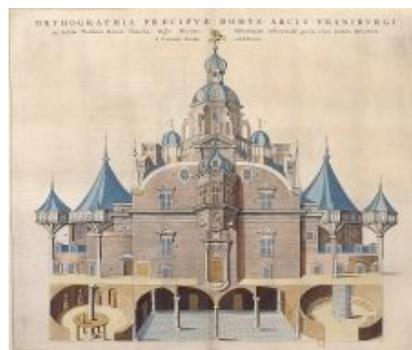

**Fig. 9.** Villa Rotonda *(left)* is a Renaissance villa just outside Vicenza, northern Italy, designed by Andrea Palladio from 1566 to 1571. It has aroused the admirations of scholars and laymen, and was a source of inspiration for the design of Tycho Brahe's manor, Uraniborg *(right)*.

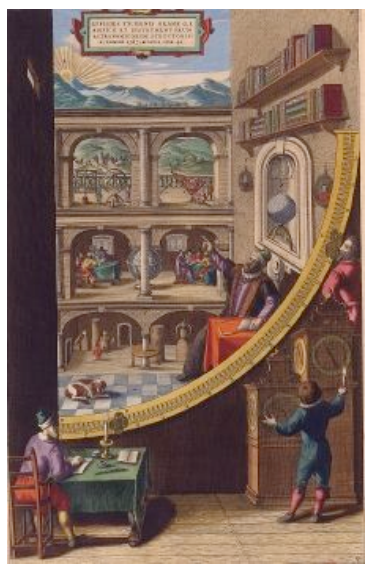

**Fig. 10.** Uraniborg (meaning « castle of the heavens ») housed an entire community dedicated to the study of the sky whose organisation reflected the extraordinary personality of its creator. As well as the observatory itself there were workshops where astronomical instruments were made, a library, a laboratory for alchemy, a paper mill and printing press, kitchen gardens, fish tanks, orchards, an irrigation system, a flour mill, etc. Uraniborg was where Tycho Brahe devoted himself to the observation of the sky. There he recorded with unprecedented accuracy the positions of the planets during 20 years.
Plate from Tycho Brahe, *Astronomiae Instauratae Mechanica*, Wandsbeck, 1598.

Johannes Kepler (1571-1630) was undoubtedly the first to integrate man's fascination with harmony into an overall vision of the world which can properly be called scientific. His avowed intention was to investigate the reasons for the number and sizes of the planets and why they moved as they did. He believed that those reasons, and consequently the secret of universal order, could be found in geometry. As he wrote later : « Geometry, which before the origin of things was coeternal with the divine mind and is God himself, supplied God with patterns for the creation of the world. »[10]

In the 1590s, when he was practising as a mathematician in Graz (then part of Styria) Kepler adopted the ideas of Copernicus. In the heliocentric model, the double motion of the earth around the sun and about its own axis explained the observed motions of the planets and stars. Kepler set out to prove that this simple hypothesis, which had been an attempt to « save the appearances », did indeed correspond with reality. In doing so, however, he noticed that the circular orbits of the planets around the sun proposed by Copernicus were not in keeping with precise observations.

In his *Mysterium Cosmographicum* (*Secret of the Cosmos)* of 1596, Kepler applied geometry in a new way to solve the problem of the relationship between the planetary orbits. Using regular polyhedra, he argued that the five « perfect » solids which Plato used to represent the five elements should correspond exactly to the intervals between the six then known planets. He went on to demonstrate mathematically how these Platonic shapes could be arranged one inside the other exactly in accordance with the structure of the solar system. For Kepler this was no coincidence : convinced that he had discovered « the secret of the cosmos », he interpreted the correspondence between the planets and the polyhedra as a novel and rational way of understanding the fundamental harmony of the universe.

Being a fervent Lutherian, Kepler wanted at the same time to glorify God, who he believed was responsible for the harmonious arrangement of the universe. This aim is stated in the very first lines of the preface to his book: « It is my intention, reader, to show in this little book that the most great and good Creator, in the creation of this moving universe and the arrangement of the heavens, looked to these five regular solids, which have been so celebrated from the time of Pythagoras and Plato down to our own, and that he fitted to the nature of those solids the number of the heavens, their proportions and the law of their motion » [11].

For Kepler, everything was an expression of harmony : poetry and music as well as mathematics and geometry. In this he can be seen as part of the great movement to reinstate mathematics as a tool for studying nature, movement which was then in its infancy but which would later recruit even more radical followers like Descartes and Galileo. It was the latter who was to declare in *The Assayer* of 1623 : « Philosophy is written in this grand book — I mean the universe — which stands continually open to our gaze, but it cannot be understood unless one fisrt learns to comprehend the language and to interpret the characters in which it is written. It is written in the language of mathematics, and its characters are triangles, circles and other geometrical figures, without which it is humanly impossible to understand a single word of it. Without these one is wandering about in a dark labyrinth. »[12]

This resurgence of mathematics would soon make numerology and numerical mysticism obsolete, just as the developement of astronomy and astrophysics were to bring about the decline of astrology as a system of interpreting the universe.

In the late 1590s, Tycho Brahe fell into disgrace after the death of his patron and left Uraniborg. He re-established himself under the patronage of the Holy Roman Emperor, Rudolf II, in Bohemia, where he invited Kepler to work with him [13]. When Brahe died in 1601, Kepler inherited the record of his uniquely detailed observations, working intensively on the information relating to the orbit of Mars. By 1605 he had determined its shape : not a circle or a combination of circles, but an ellipse with the sun at one focus [14]. Thus, in his *Astronomia Nova* (*The New Astronomony*)[15] of 1609 Kepler modified the Copernican model of the solar system, shattering the belief in the perfection of cicular orbits ; he event went as far as suggesting that it is the sun's influence which causes the planets to follow their orbits. Copernicus did not ascribe any physical superiority to the sun, nor, more significantly, any real dynamic influence. Kepler was the first to assign a predominant role to the sun, to which he attributed a « motivic force », an abstract entity which clearly anticipated the Newtonian theory of universal attraction.

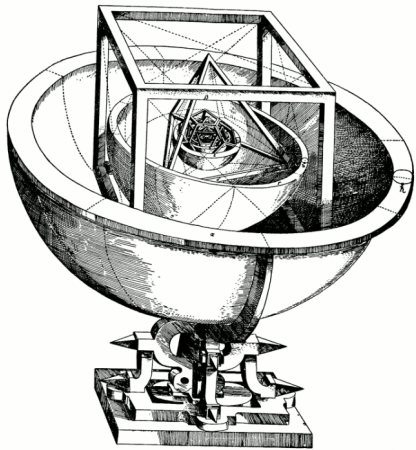

**Fig. 11.** This famous engraving, which Kepler made at the age of 25, depicts his view of the universe based on the appication of regular polyhedra. He inserted an octahedron between the orbits of Mercury and Venus, an icosahedron between Venus and Earth, a dodecahedron between Earth and Mars, a tetrahedron between Mars and Jupiter, and a cube between Jupiter and Saturn. Extensive calculation led Kepler to the view that « symmetrical shapes can be placed so precisely inside each other to separate the relevant orbits that if a layman were to ask how the heavens are supported to prevent them from falling, the answer would be simple. » Not content with this explanation of celestial harmony, Kepler had the idea of developing it into an artefact and devised a « cosmic bowl » which would actually dispense beverages representing the symmetry of the polyhedra and the harmony of the planets. In February 1596 he went to visit his patron, Duke Frederick of Würtemberg, to ask him to authorise the construction of a model of the universe in the form of a bowl. The planets would be made of precious stones – a diamond for Saturn, a sapphire for Jupiter, a pearl for the moon, etc.- and the beverages contained in each planetary sphere would be fed through invisible pipes to seven taps around the rim of he bowl. Mercury would provide brandy, Venus mead, Mars a strong vermouth, and so on. The project was too costly and was never realised.
Plate from *Mysterium Cosmographicum*, Tübungen, 1596.

In *Harmonices Mundi (Harmony of the World,* 1619), Kepler finalised his laws of planetary motion, which are still valid today. The third of these, known as the law of orbital periods, establishes a « harmonic » relationship, in the mathemacial sense of the word, between the axis of the orbit and the orbital period, and enables the movement of the planets to be described much more easily.

Although Kepler was forced to abandon the regular polyhedra for describing the structure of the solar system, he retained his fascination for these almost perfect shapes. In looking at the group of semi-regular poyhedra (rhomboids, prisms, etc.) which incorporates the group of regular solids but also includes non-convex shapes, Kepler discovered « stellation ». In his *De Niva Sexangula* (*The Six-Cornered Snowflake*) of 1610, he remarked on the hexagonal shape of snowflakes and made the connection with the shape of certain crystals [16]. He accounted for this by premonitory geometric reasoning, by asking himself, for example, how to pack regular solids in the most compact possible way. Nature produces these types of packings: pomegranate seeds, of rhombohedric shape, occupy the least possible space in the fruit, and bee honeycombs have a hexagonal shape that allows them to contain the greatest amount of honey. But Kepler above all caught sight of the underlying symmetry principles that preside in the ordering of the world, on all distance scales, from crystals to planetary orbits and the cosmos as a whole. He believed that geometric symmetry was the natural language with which God expressed himself in Creation.

Kepler wanted to do more than create a simple model or describe the results of his experiments and observations ; he wanted to explain the causes of what he saw. At this turning point between ancient and modern thinking Kepler was steeped in a tradition which connected cosmology explicitly with the notion of divine harmony. But what Kepler sought to express was not the numerical myticism of the Pythagoreans ; his starting point was geometric patterns, which he saw as « logical elements ». This makes him one of the greatest innovators in the history of science.

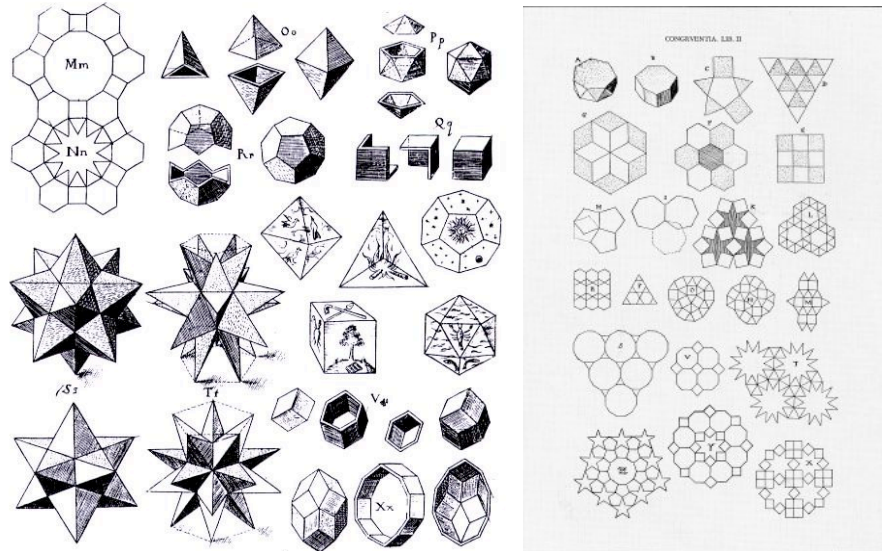

**Fig. 12.** Kepler believed that regular polygons held the secret of the origin of the cosmos. From this assumption he derived the « three most perfect congruencies » (in other words mosaics or, in modern mathematical parlance, tessellations) derived from one of the three regular polygons (triangle, square or hexagon) and « nine perfect congruencies » derived from a combination of two or three of them, as well as an indeterminate number of other polygons and star shapes. The construction of mosaics is integral to Kepler's thinking and still relevant today to non-Euclidean geometry and topology. Plates from *Harmonices Mundi*, Linz, 1619.

In succeeding centuries, many artists have drawn polyedra (figs. 13, 14), including Mauritz Cornelius Escher (1898-1972) – whose brother was a professor of crystallography at Leiden University. Salvador Dali (1904-1989), in particular, painted the impressive *Sacrament of the Last Supper* (1955), which takes place in a room of dodecahedral shape, surrounded with large pentagonal bay windows (fig. 15). His other painting, *In search of the fourth dimension* (1979), displays two dodecahedra, one solid, the other one airy and fiery (fig. 16).

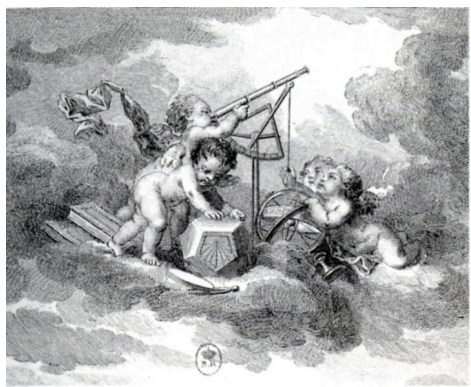

**Fig. 13.** This anonymous 17th century engraving shows cherubs playing in the sky with astronomical instruments and a dodecahedron, symbolising Aristotle's fifth element, ether.

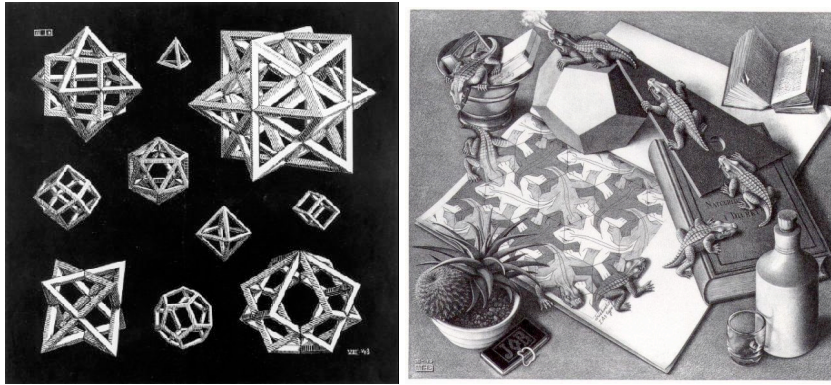

**Fig. 14.** *Left :* M.C. Escher, *Study of Stars,* wood engraving (1948). *Right :* a dodecahedron sits on the table in Escher's lithograph print *Reptiles* (1943).

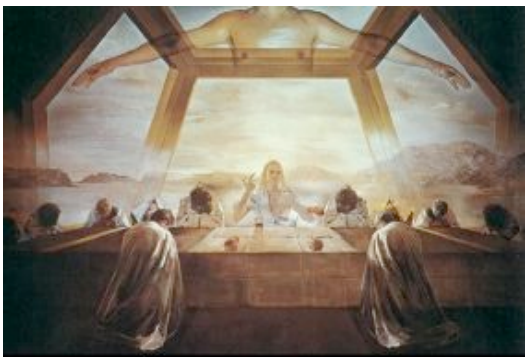

**Fig. 15.** Salvador Dali explicitly used the golden ratio in his masterpiece, *The Sacrament of the Last Supper*. The dimensions of the canvas are a golden rectangle. A huge dodecahedron, with edges in golden ratio to one another, is suspended above and behind Jesus and dominates the composition.

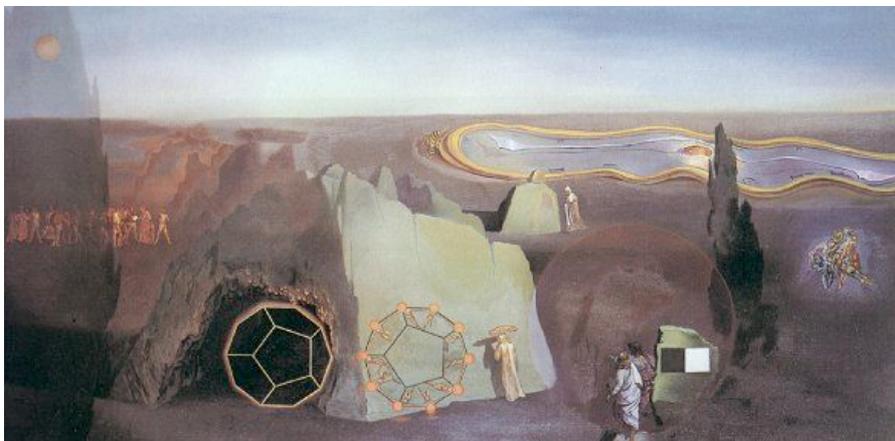

**Fig. 16.** Salvador Dali, *In search of the fourth dimension* (1979), oil on canvas, Fondacio Gala-Salvador Dali, Figueres. This painting exceptionally captures many interpretations of the « fourth dimension ». It depicts a melting, wavering clock which appears to be either resting on the ground or floating. One is unable to truly tell, bending the ideas of perception of air and solid ground, as well as questioning the reality of the relationship of the two on Earth. This question is once again raised on the left side of the painting in which a group of people appear to be gathering in a circle, suspended in the middle of the air. Besides the very foreground of the painting, Dali purposely confuses the idea of solid ground and the air above, allowing one to develop one's own elucidation of the relationship.

It is above all in in crystallography, in cosmology and in topology that the polyhedra have revealed all of their explanatory potential. Modern relativistic cosmology investigates the possibility that space itself is in some way polyhedral and that the cosmos as a whole has a crystallographic structure [17]. According to the theory of general relativity, space has a geometric structure characterised by curvature and topology. This idea fascinated several of the founders of 20<sup>th</sup> century cosmology such as De Sitter, Friedmann and Lemaître ; it then rather lost favour before regaining popularity in recent years. In topologically « wraparound » universe models [18], space is represented by a fundamental polyhedron. The simplest of these models use cubes or parallelepipeds to create a toroidal space, but there is an infinite number of variations. The common feature of these fundamental polyhedra is that they have some symmetries (realised by a so-called holonomy group), so that one face can be related to another ; the corresponding points on each face are therefore « linked » in such a way that physical space is the result of a complex folding process. The fundamental geometric symmetry of the universe is matched by the symmetry of the polyhedron.

From the point of view of the celestial observer, these wraparound universe models introduce a radically new perspective. The usual interpretation of the sky is that it consists of a straightforward projection of the space, which is vast if not infinite : each point of light that we can see corrsponds to a specific star, galaxy or other celestial body – the further always the fainter. This is not at all the case with a wraparound model, according to which some celestial bodies are represented by a whole series of « ghosts » so that what we see in the sky is not the universe as it really is, but several different images of the universe, from different angles and distances, superimposed upon one another (fig. 17). Cosmic crystallography methods [19] have been developed to look at the 3-dimensional observed distribution of high redshift sources (e.g. galaxy clusters, quasars) in order to unveil specific correlations which would signal repetitions of patterns, much like those observed in crystals.

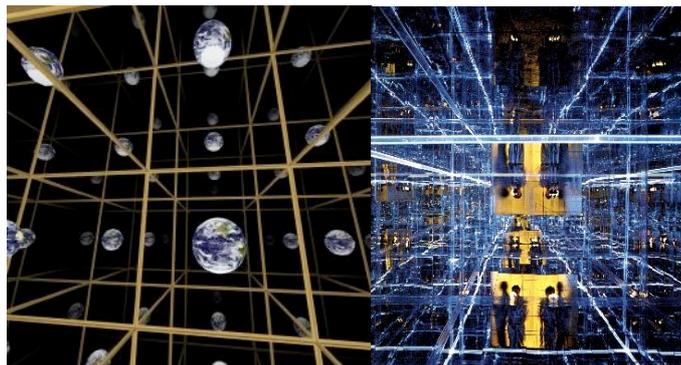

**Fig. 17.** *Left :* By analogy with the two-dimensional case, the three-dimensional hypertorus is obtained by identifying the opposite faces of a parallelepiped. The resulting volume is finite. Let us imagine a light source at our position, immersed in such a structure. Light emitted backwards crosses the face of the parallelepiped behind us and reappears on the opposite face in front of us; therefore, looking forward we can see our back. Similarly, we see in our right our left profile, or upwards the bottom of our feet. In fact, for light emitted isotropically, and for an arbitrarily large time to wait, we could observe ghost images of any object (here the Earth) viewed arbitrarily close to any angle. The resulting visual effect would be comparable (although not identical) to what could be seen from inside a parallelepiped of which the internal faces are covered with mirrors. Thus one would have the visual impression of infinite space, although the real space is closed (courtesy Jeff Weeks). *Right :* The architect Serge Salat has designed astonishing installations where the spectator loses all spatial landmarks, and has the impression of entering a crystalline space that repeats out to infinity in every direction [20].

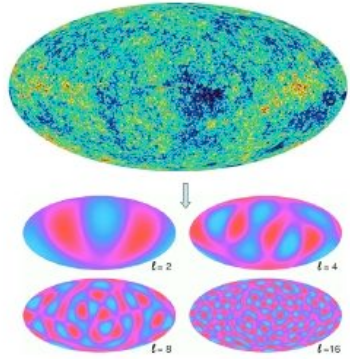

**Fig 18**. The Cosmic Microwave Background harmonic decomposition from WMAP observations.

Cosmologists may also study the topology of space by analyzing in great details the temperature fluctuations of the Cosmic Microwave Background radiation (CMB). Like acoustic waves, the CMB temperature fluctuations can be decomposed into a sum of spherical harmonics (fig. 18). The first observable harmonics is the quadrupole (whose wavenumber is l=2). The 2003-2008 data collected by the WMAP satellite, which produced a high resolution map of the CMB, showed that the temperature fluctuations on small and mean scales (i.e. concerning regions of the sky of relatively modest size) are compatible with the infinite flat space hypothesis. However, on angular scales of about 90°, corresponding to the quadrupole harmonics, the observed correlations are notably weaker that those predicted by the standard model.

The unusually low quadrupole value means that long wavelengths are missing, may be because space is not big enough to sustain them. Such a situation may be compared to a vibraring string fixed at its two extremities, for which the maximum wavelength of an oscillation is twice the string length. A possible explanation of such a phenomenon relies on a model of finite, wraparound space whose size constrains the wavelengths below a maximum value. The present author and his collaborators [21] proposed the Poincaré dodecahedral space as a best fit model (fig. 19). Its volume is 120 times smaller than that of the hypersphere with the same radius of curvature, which gives rise to a fascinating topological mirage (fig. 20).

To convey our vision, we illustrated a beguiling sphere of curved pentagons and a dense view through the surface of a hypersphere tiled with 120 spherical dodecahedra. Such a visualization of a finite universe, albeit one from which we can exit through one face and simultaneously enter through the opposite one, relies upon a keplerian form of mental sculpture that may be described as plastic as well as algebraic.

Note that the symmetries of the dodecahedron play also a role in quantum mechanics. Roger Penrose gave an ingenious proof of Bell's nonlocality theorem using a special set of states of a spin(-3/2) particle ; these states have a direct connection with the geometry of a dodecahedron. A popular account of the Penrose dodecahedron, presented as a puzzle to whet the appetite of the intelligent layman in the mysteries of the quantum theory, was given in [22].

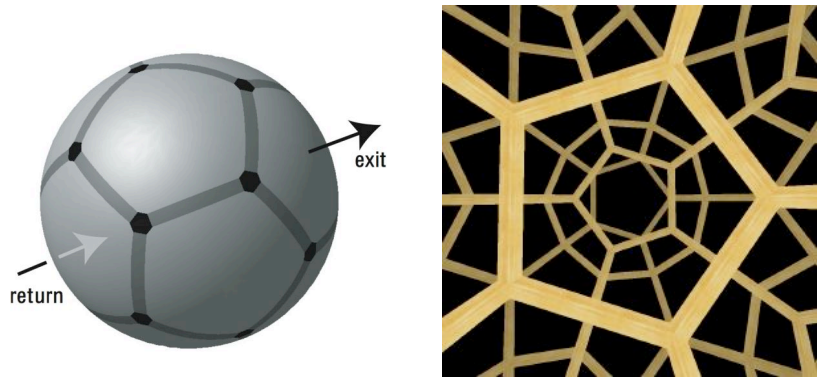

**Fig. 19.** *Left :* Poincaré Dodecahedral Space can be described as the interior of a spherical dodecahedron such that when one goes out from a pentagonal face, one comes back immediately inside the space from the opposite face, after a 36° rotation in the clockwise direction around the axis orthogonal to the face. Such a space is finite, although without edges or boundaries, so that one can indefinitely travel within it. *Right :* View from inside PDS perpendicularly to one pentagonal face. In such a direction, ten dodecahedra tile together with a 1/10th turn to tessellate the universal covering space $S^3$. Since the dodecahedron has 12 faces, 120 dodecahedra are necessary to tessellate the full hypersphere.

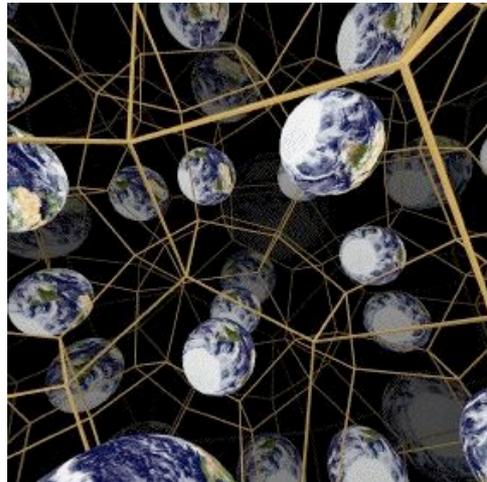

**Fig. 20**. An observer living in PDS has the illusion to live in a space 120 times vaster, made of tiled dodecahedra which duplicate like in a mirror hall. From CurvedSpaces program, courtesy J. Weeks.

## 2. The art of tessellations

Imagine a room paneled with mirrors on all four vertical walls, and place ourselves somewhere within the room: a kaleidoscopic effect will be produced in the closest corner. Moreover, the repeated reflections of each pair of opposing mirrors ceaselessly reproduce the effect, creating the illusion of an infinite network extending in a plane. This tiling of an infinite plane by a repeating pattern is called a *tessellation* (*tessella* being the name for a mosaic tile) of the Euclidean plane. It can be generalized to n-dimensional spaces. As we have seen above, the ordinary 3D-Euclidean space can be tessellated by an infinite number of hypertori, while the finite hypersphere can be tessellated by 120 Poincaré dodecahedral spaces.

Tessellations are integral to the topological classification of manifolds. For instance, take a rectangle, called a fundamental domain, and glue its opposite edges two by two. We obtain a flat torus, a surface whose topology is identical to that of a ring but whose curvature is everywhere zero. The mathematical transformations used to glue the edges (here translations) form a group of symmetries, called the holonomy group. Starting from the rectangle and acting with the transformations of the holonomy group on each point, one creates a number of replicas of the rectangle : we produce a tessellation of a larger space (here the Euclidean plane), called the universal covering space. The latter can be thought of as an unwrapping of the original rectangle. The topology of a space is thus entirely specified if one is given a fundamental domain, a particular group of symmetries (the holonomy group), and a universal covering space that is tessellated by fundamental domains.

In 1891, the Russian mineralogist Y. S. Fedorov demonstrated that the number of symmetry groups allowing one to regularly tessellate the plane is equal to 17. In 1922, the archeologist Andreas Speiser remarked that these 17 groups were discovered empirically 4000 years earlier in decorative art. While studying Greek weavings, the pavings of Egyptian temples, and the mosaics of the Alhambra in Granada, Spain, he noticed that they were composed of identical patterns, combined by simple or composite symmetries, with all possible operations reducing to the 17 groups identified by Fedorov. The innumerable variety of planar decorations can therefore be reduced to an exhaustive mathematical description.

The freedom of choice in selecting the fundamental cell allows one to tile the entire plane with a pattern of arbitrary shape. This precept was applied by the Dutch engraver M.- C. Escher. In 1936, the young artist traveled to the Alhambra of Granada, where he was fascinated by the Moorish tilings. Soon after this visit, he read a popularizing article that the Hungarian mathematician George Polya had published in 1924 on the symmetry groups of the plane. Without understanding the abstract aspect, Escher was able to extract the 17 symmetry groups that were described there. Between 1936 and 1941, he applied his new knowledge in an impressive series of engravings presenting all possible periodic tilings. Taking the opposite approach from Islamic art, which had to confine itself to purely geometric patterns, Escher used animal or human forms: butterflies, birds, fish, lizards, and imps (fig. 21). He entered into contact with renowned mathematicians like Donald Coxeter and Roger Penrose, and worked in collaboration with them. By introducing colored patterns into his engravings — a supplementary parameter that was not taken into account in Fedorov's classification — Escher opened up a new field of geometry, the theory of polychromatic symmetry groups, subsequently studied by Coxeter [23].

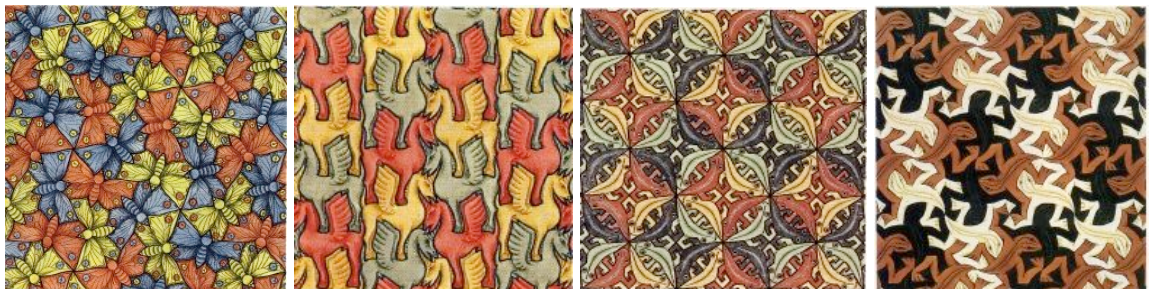

**Fig. 21.** Four Escher tilings of the Euclidean plane using animal patterns.

Let us now take two tori and glue them to form a « double torus ». As far as its topological properties are concerned, this surface with two holes can be represented as an eight-sided polygon (an octagon), which can be understood intuitively by the fact that each torus was represented by a quadrilateral. But this surface is not capable of tessellating the Euclidean plane, for an obvious reason: if one tries to add a flat octagon to each of its edges, the eight octagons will overlap each other. One must curve in the sides and narrow the angles, in other words pass to a hyperbolic space: only there does one succeed in fitting eight octagons around the central octagon, and starting from each of the new octagons one can construct eight others, *ad infinitum*. By this process one tessellates the Lobachevsky hyperbolic plane (fig. 22).

A fascinating representation of a hyperbolic tessellation was given by Poincaré. A conformal change of coordinates allows us to bring infinity to a finite distance, with the result that the entire Lobachevsky space is contained in the interior of the unit disk. Escher created a series of prints entitled *Circle Limit*, in which he used Poincaré's representation (fig. 23).

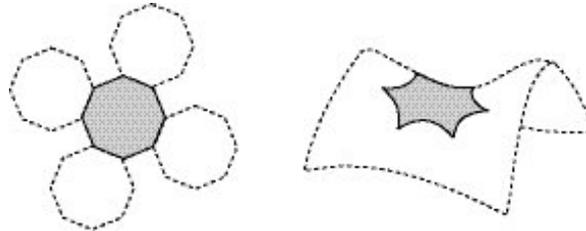

**Fig. 22.** While it is impossible to tessellate the Euclidean plane with octagons, the hyperbolic plane can be tiled by octagons cut from the hollow of a saddle. The eight corners of the octagon must all be identified as a single point; this is the reason why one must use a negatively curved octagon with angles of 45 degrees (8x45 = 360), in place of a flat octagon, whose angles are each 135 degrees.

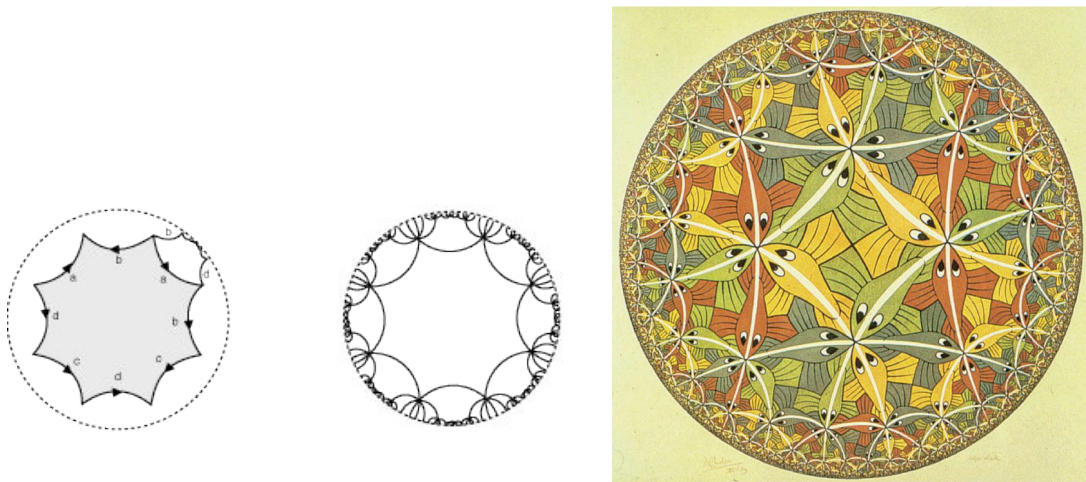

**Fig. 23.** *Left* : Poincaré's Representation of the Hyperbolic Plane. By acting with the holonomy group on each point of the fundamental octagon, and repeating the process again and again, one creates a tessellation of the hyperbolic plane by regular and identical octagons. Poincaré demonstrated that the hyperbolic plane, normally infinite, could be represented entirely within the interior of a disk, whose edge represents infinity. Poincaré's model deforms distances and shapes, which explains why the octagons seem irregular and increasingly tiny as we approach the boundary of the disk. All of the lines in the figure represent straight lines of the hyperbolic plane, and meet the boundary at a right angle. *Right* : In this 1959 woodcutting entitled Circle Limit III, Escher has used the representation given by Poincaré to pave the hyperbolic plane using fish.

Symmetry is one of the most fundamental concepts in geometry, whose principal concern is to find « pure » shapes – the equivalent of the physicist's search for fundamental elements. It is so prevalent in nature, from the human body to crystals, atoms, particle physics or cosmology that it is difficult to imagine it not being central to our understanding of the world. Although symmetry was studied by the French mathematician Evariste Galois in the early 1830's and by the German Emmy Noether around 1916, its importance was not fully understood until the developement of group theory later in the 20[th] century.

Symmetry is also essential in the arts. The (subjective) notion of beauty is, however, often associated with a slight asymmetry. The most beautiful faces are not exactly symmetrical ; the best architects mix the symmetrical with the unexpected. Similarly, physicists study symmetry breakdowns and show how they are as fundamental to nature as symmetry itself.

Whereas perfect symmetry is static, broken symmetry introduces dynamics (e.g. in particle physics, phase transitions, high-energy cosmology, etc.). Following these lines of reasoning, I tried to transpose the concept of dynamical broken symmetry in various artworks.

I chose Indian ink and different engraving techiques to express my fascination for the darkness of the universe and the invisible architecture of space-time. By repeating ceaselessly black and white checkerboards, I intend to evoke the notion of infinity and generate the dizziness of the glance. By disrupting the laws of classical perspective, I conjugate architectural immobility and time unsettledness. Elements are immersed in a black ocean, and such an interaction is aimed to give the sensation of perpetual time. Infinity is that aspiration felt by the man held on ground by gravitation. I tried to depict this sensation by creating vanishing points towards which one feels irresistibly attracted.

In *Black Hole* (1979), an optical illusion (the checkerboard ground seems to be negatively curved, although lines are perfectly straight) is created by a violation of the usual laws of perspective : besides the three vanishing points $P_1$-$P_2$-$P_3$ on the eye-level horizon line to which parallel lines converge, the adjonction of an additional point $P_4$ on the vertical attracts the glance and creates an apparent bending of the checkerboard towards the bottom of the hole (fig. 24). The lithograph *Big Bang* (1992) exploits the spatial vocabulary of perspective to evoke realms beyond the three-dimensional. Whereas Escher relied on contradictions and oscillating ambiguity in his graphic art, I tried to suggest plunging, interpenetrating and vertiginous illogicalities of dynamic space. Spewing from the vertex in the top left-hand corner, matter organizes itself into structures on the right; the tumbling dice on the left imply irreversible disorganization arising from chance (fig. 25).

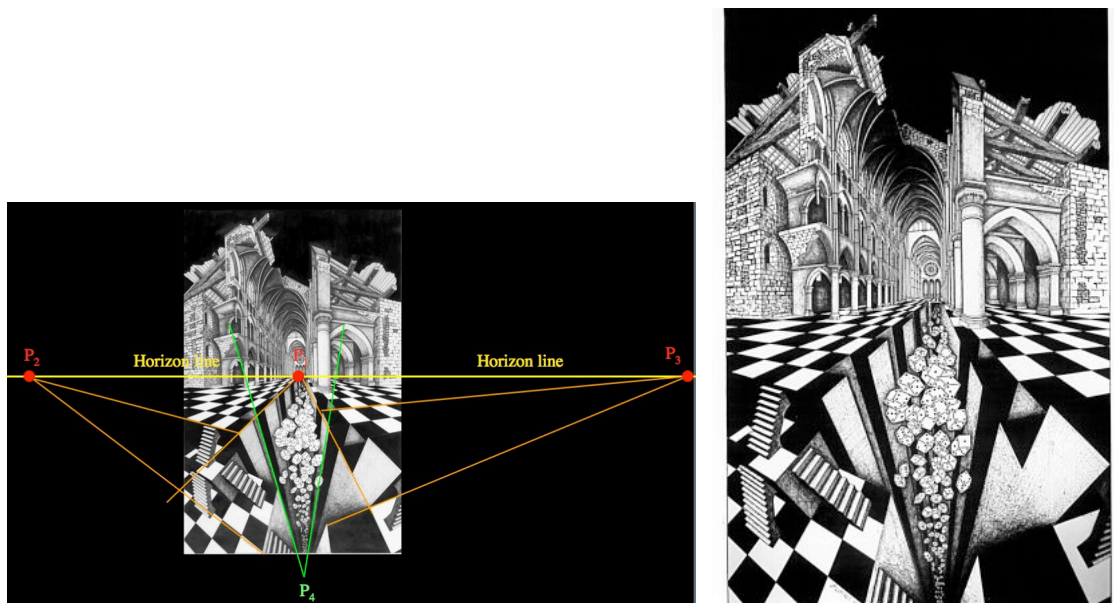

**Fig. 24.** J.-P. Luminet, Black Hole, Indian ink, 1979 (private collection)

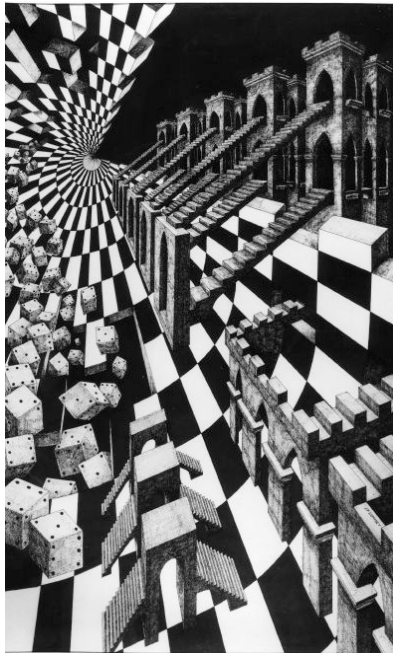

**Fig. 25**. J.-P. Luminet, *Big Bang*, lithograph print (1992)

Tilings 1 and 2 stem directly from considerations on topology and tessellations. As we have seen above, repeating a square by a two dimensional group of translations generates a perfectly homogeneous space, which reflects the topological properties of the torus (fig. 26). Now, repeating a square by a less symmetrical holonomy group, generated by a translation and a translation composed with a flip, generates a non homogeneous universal covering space, which reflects the topological properties of a Klein bottle (a closed, non orientable Euclidean surface). Transposed to an artwork, such a « broken translational symmetry », obtained by repeating a given pattern according to the rules of the Klein bottle rather than the rules for a torus, amazingly creates a dynamical effect (fig. 27).

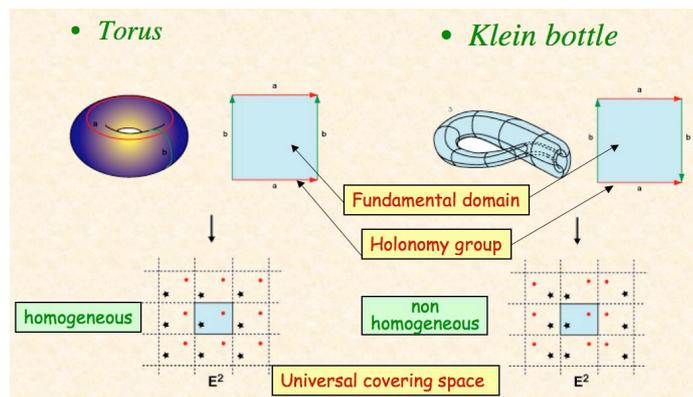

**Fig. 26.** Homogenous vs. non homogeneous space

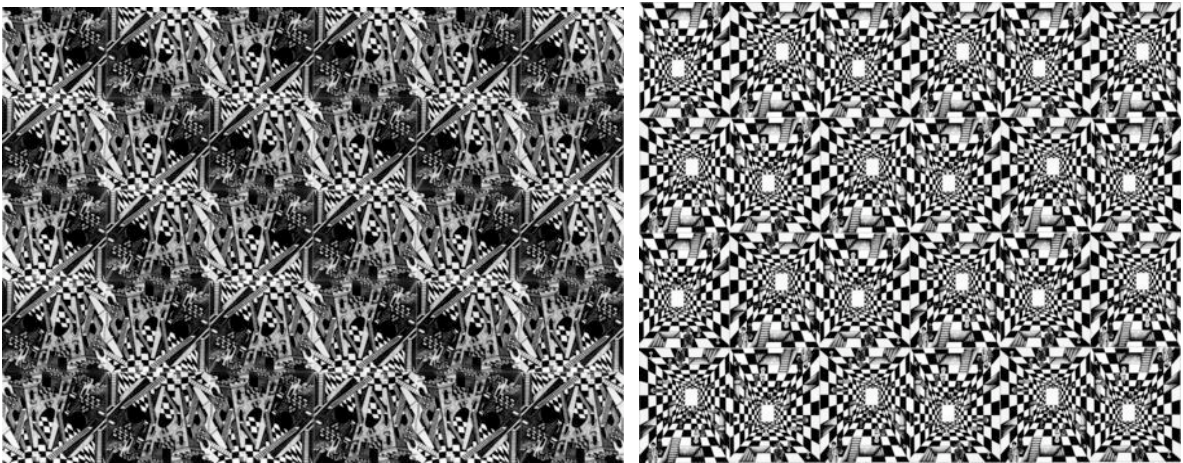
**Fig. 27.** J.-P. Luminet, *Tiling 1 and Tiling 2,* collages (2005), Collegio Cairoli and University of Pavia, Italy.

### 3. Music of the spheres

Plato's *Republic* is the first written reference to the harmony of the celestial spheres. Even earlier than this, Pythagoras had asserted that the entire universe echoed with harmonious singing. The regular proportions of the world were believed to correspond to musical intervals. In ancient times music was considered to be a branch of mathematics. It was one of the four major disciplines (the *quadrivium*) alongside with geometry, arithmetics and astronomy. In the Middle Ages the Italian theorist and composer Franchino Gafori acquired a Latin translation of Ptolemy's *Harmonics* and proceeded to write his own treatise on *The Practice of Music* (1496), whose frontispiece is a synthesis of contemporary theories about the music of the spheres. In 1572 the Italian humanist and music historian Girolamo Mei wrote to the Florentine composer Vincenzo Galilei (Galileo's father) on the subject of Greek music with particular reference to the power it might have over the emotions. Galilei and his colleagues developed these ideas into a new musical aesthetic which culminated in Monteverdi's *L'Orfeo* of 1607, the first « dramatic opera ». But Kepler, in his *Harmony of the World*, took the idea of a link between music and astronomy a step further. Having abandoned the use of polyhedra to explain the planetary orbits, which he had discovered were elliptical, Kepler saw that this discovery could lead to even greater harmony. Compared with a circle, an ellipse had an additional parameter, eccentricity, which was defined by the relationship between its long and its short axes. Many centuries earlier the Pythagoreans had imagined that the faster a planet travelled along its orbit the higher the pitch of the sound it emitted, and that celestial harmony was the product of the different sounds emitted by all celestial bodies. Kepler attempted to revive this idea in the context of his discovery of elliptical orbits, which meant that the speed of the planets varied according to the degree of eccentricity of their orbit; they did not each produce a single note, as the anciens believed, but a range of notes determined by their maximum and minimum speeds. Kepler associated each planet with two or more integer number ratios and related these to musical intervals (e.g. the ratio 4:3 is related to the interval of a fourth, 3:2 to a fifth and 2:1 to an octave) to create an entire harmonic sructure, which he saw as a reflection of the structure of the solar sysem itself : « the multiple reflections from the heavenly bodies create a melody, which is the music to which the sublunary world dances » – to which he added : « The Earth sings the notes C, D, C, from which one might conjecture that our world is prevailed upon by Calamity and Disaster. »

Kepler's ideas were as much intuitive and psychological as logical. They may seem naïve today, yet the music of the cosmos has never ceased to inspire poets and composers in one way or another.

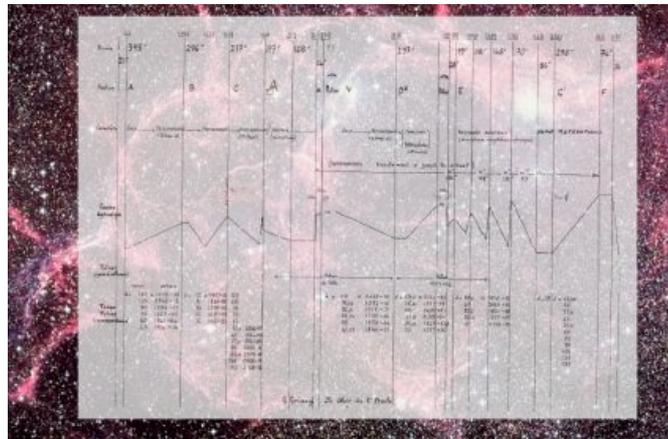

**Fig 28.** This unusual piece of music incorporates the rhythms generated by pulsars – rapidly rotating neutron stars. The tempi and rhythms of the piece are entirely derived from the speed of rotation and the frequencies of theses remnants of stars, which have exploded in supernovae. The date and time of the concert are determined by the times when the radio waves emitted by the pulsars can be detected, since they are picked up by a huge radiotelescope and transmitted directly to the concert hall. *Background image :* one such pulsar is the Vela pulsar, which rotates 11 times a second amid the debris of a supernova which exploded 12,000 years ago. G. Grisey and J.-P. Luminet, *Le Noir de l'étoile*, for six percussionists, tape and retransmision in situ of astronomical signals, 1991. Ed. Ricordi, Milano.

In 1991, I collaborated with the Frenchman Gérard Grisey to conceive a « cosmic piece », *Le Noir de l'Etoile*, for six percussionists and « guest stars » : pulsars sending their unchanged rhythms from the depths of space, like celestial metronomes guiding the hands of the performers here on earth. The score requires the six players to surround the audience, and in addition to the pre-recorded tape there should be live signal transmissions as well. While it is clear where percussion ends and pulsars begin, their overlapping moments beg deciphering the random beats from space and perceiving the man-made sounds as organic entities.

Our society today is prey to may questions regarding its future, and *Le Noir de l'Étoile* is perhaps one of the responses to a more universal world. These sounds coming from the stars, carrying in them the testimony of a past world, can be compared to a kind of voice of rediscovered wisdom, a proposal for a language that knows no barriers, whether they be cultural, religious or territorial. In that respect, this work remains utopic.

### 4. Conclusion

Unlike art, science is an essentially rational discipline, but this is not to say that beauty has no place in it. Thoughout history scientists have been guided by aesthetic principles and 20th century physicists considered the relationship of aesthetics to the construction of matter and to the choice between theoretical models. The English theoretical physicist Paul Dirac stated [24], « It seems that if one is working from the point of view of getting beauty in one's equations, and if one has really a sound insight, one is on a sure line of progress. » His famous equation appeared to contradict the facts, yet it became the foundation of modern field theory and is now universally accepted.

It is well known that hree of the many characteristics of quarks have been called beauty, charm and truth. But although it is acknowledged that a theory should be judged by its beauty, it is less clear how that beauty should manifest itself : as harmonious geometric shapes, as symmetrical mathematical equations, as simplicity or clarity… Various criteria have been suggested and different ones favoured at different times, since this is probably the most subjective area in all scientific research and the one most affected by social and cultural factors. Whereas men like Kepler and Einstein made absolute claims, modern physicists such as John Wheeler and Roger Penrose insist on the significance of aesthetics in choosing and evaluating scientific theories. According to Penrose, « It is a mysterious thing, in fact, how something which looks attractive may have a better chance of being true than something which looks ugly.» [25]

Today many physicists are pinning their hopes for an explanation of the world's fundamental structure on superstring theory (see e.g. [26]). But this idea, although now more than 30 years old, has yet to generate a single testable prediction, let alone be substantiated. Its attraction is undoubtedly its elegance and the many symmetrical features it entails. Time will tell whether there are indications of its veracity.

Creativity in art and science is integral to recreate a modern « humanism of knowledge », according to which the arts and sciences are not to be conflated because they work in very different ways, with illogical and logical means, but they well up from the same instincts and intuitions. Like science, the arts have developed some fundamental rules: poets have established a series of set forms; painters have pursued variations on perspective; and musicians have developed a harmonic language that is at once simple and intricate. The modes of formal organization have varied over time, yet a work of art habitually relies on tension between set structure and unpredictable freedom, regardless of the kind of formal organization used.

It is also right to acknowledge the passionate engagement that potentially flows undetected beneath the dry crust of the standard scientific paper. Knowledge must not be separated from emotion ; their commont root is amazement about the world, which is expressed through a harmonious integration of all those intellectual and creative faculties that we use to respond to the wonder of things, both immense and minute.

As said Albert Einstein, « man tries to make for himself in the fashion that suits him best a simplified and intelligible picture of the world; he then tries to some extent to substitute this cosmos of his for the world of experience, and thus to overcome it. This is what the painter, the poet, the speculative philosopher, and the natural scientists do, each in his own fashion. »[27] Geometric imagination is a pretty efficient one.